\documentclass[twocolumn,pre,showpacs]{revtex4}
\usepackage{graphicx}

\begin{document}

\title{Optimized ensemble Monte Carlo simulations of dense Lennard-Jones fluids}

\author{Simon Trebst$^{1,2}$,  Emanuel Gull$^{2}$, Matthias Troyer$^{2}$}

\affiliation{$^{(1)}$Computational Laboratory,  Eidgen\"ossische Technische Hochschule
 Z\"urich, CH-8092 Z\"urich, Switzerland}
\affiliation{$^{(2)}$Theoretische Physik, Eidgen\"ossische Technische Hochschule
 Z\"urich, CH-8093 Z\"urich, Switzerland}
\date{\today}

\begin{abstract}
We apply the recently developed adaptive ensemble optimization technique to simulate dense Lennard-Jones fluids and a particle-solvent model by broad-histogram Monte Carlo techniques.
Equilibration of the simulated fluid is improved by sampling an optimized histogram in radial coordinates that shifts statistical weight towards the entropic barriers between the shells of the liquid.
Interstitial states in the vicinity of these barriers are identified with unprecedented accuracy by sharp signatures in the quickly converging histogram and measurements of the local diffusivity.
The radial distribution function and potential of mean force are calculated to high precision.
\end{abstract}

\pacs{05.10.Ln,61.20.Ja,02.70.Uu}

\maketitle


The free energy landscapes of complex systems are characterized by many local minima that are separated by entropic barriers. The simulation of such systems with conventional Monte Carlo \cite{MonteCarlo} or molecular simulation \cite{MolecularSimulation} methods is slowed down by long relaxation times due to the suppressed tunneling through these barriers.
Extended ensemble simulations address this problem by broadening the range of phase space which is sampled in the respective reaction coordinate. Recently, an adaptive algorithm \cite{OptimalEnsemble} was introduced that explores entropic barriers by sampling a broad histogram in a reaction coordinate and iteratively optimizes the simulated statistical ensemble defined in the reaction coordinate to speed up equilibration. The key idea of the approach is to measure the local diffusivity along the reaction coordinate, thereby identifying the bottlenecks in the simulations and then using this information to systematically shift statistical weights towards these bottlenecks in a feedback loop. The so optimized histogram converges to a characteristic form exhibiting peaks at the bottlenecks of the simulation, e.g. in the vicinity of the entropic barriers.
The simulation of an optimized ensemble results in equilibration times which can be substantially lower compared to other extended ensemble simulations that aim at sampling a flat histogram in the respective reaction coordinate \cite{OptimalEnsemble,SlowDown}. Flat-histogram algorithms include the multicanonical method \cite{Multicanonical}, simulated and parallel tempering \cite{Tempering}, broad histograms \cite{BroadHistograms} and transition matrix Monte Carlo \cite{TransitionMatrixMC} when combined with entropic sampling, as well as the adaptive algorithm of Wang and Landau \cite{WangLandau} and subsequent improvements \cite{WL-Improvements} and variations \cite{WL-Variations} thereof.

\begin{figure}[b]
  \includegraphics[width=96mm]{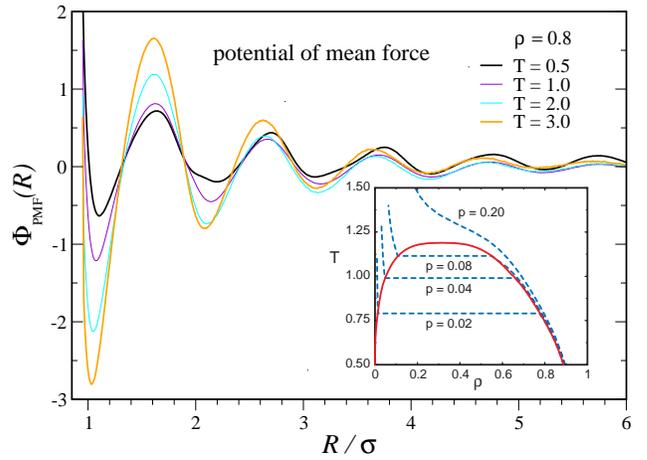}
  \caption{Potential of mean force for a Lennard-Jones fluid at density $\rho=0.8$
                  for varying temperatures.
                  The inset shows the vapor-liquid equilibrium (solid line) of a Lennard-Jones fluid
                  calculated by an extended ensemble simulation in Ref.~\cite{Yan:02}.  }
  \label{Fig:PMF}
\end{figure}

In this paper we demonstrate how the ensemble optimization method can be applied to a system that exhibits multiple barriers along a continuous reaction coordinate. 
Specifically, we simulate dense Lennard-Jones (LJ) fluids in two dimensions with a pair-wise potential of the form
\begin{equation}
   \Phi_{\text{LJ}}(R) = 4\epsilon 
      \left[ \left(\frac{\sigma}{R}\right)^{12} -  \left(\frac{\sigma}{R}\right)^6 \right] \;,
   \label{Eq:LJ}
\end{equation}
where $\epsilon$ is the interaction strength and $\sigma$ a length parameter.
In the fluid the interaction between two particles at distance $R$ in the presence of many surrounding particles can be described by a potential of mean force (PMF) that has the form
\begin{equation}
  \Phi_{\text{PMF}}(R) = - \frac{1}{\beta} \ln g(R) \;,
\end{equation}
where $g(R)$ is the radial pair distribution function.
The PMF and radial pair distribution function show a sinusoidal modulation revealing shell structures that form in the liquid as illustrated for various temperatures in Fig.~\ref{Fig:PMF}. Between the shells multiple entropic barriers form which confine particles to the shells and thereby suppress thermal equilibration between the shells.
Over the last decades Lennard-Jones fluids have been extensively investigated by numerical simulations \cite{MonteCarlo,MolecularSimulation} and provide an ideal test system to study novel algorithms. Extended ensemble simulations of dense Lennard-Jones fluids have been performed in energy space  using Wang-Landau sampling \cite{Yan:02,Shell:02} and  in radial coordinates using an adaptive integration method (AIM) \cite{AIM}.
Here we determine the optimized ensemble in radial coordinates and its characteristic histogram for various temperatures and subsequently calculate the radial distribution function and potential of mean force. Approaching vapor-liquid coexistence we observe distinct features in the optimized histogram that allow us to identify interstitial states in the vicinity of the entropic barriers between the shells of the liquid.

{\em Ensemble optimization --}
In our Monte Carlo simulations the system does a Markovian random walk in configuration space which is described by particle positions in real space. This random walk is projected onto a random walk in two reaction coordinates, namely onto the energy $E$ of a configuration which can be calculated by evaluating the pair-wise potential in Eq.~(\ref{Eq:LJ}) as well as onto the radial distance $R$ of two arbitrarily chosen particles in the system.   
In energy space we simulate a canonical ensemble with weights $w(E) = \exp(-\beta E)$, where we fix the inverse temperature $\beta$. For the second reaction coordinate, the radial distance $R$ between two particles, we define an additional broad-histogram ensemble with weights $w(R)$ as explained below. 
These two ensembles define the acceptance probabilities for moves based on the Metropolis scheme
\begin{equation}
  p_{\text{acc}}(R_1, E_1 \rightarrow R_2, E_2) 
    = \min \left( 1, \frac{w(R_2)}{w(R_1)} \cdot e^{-\beta (E_2-E_1)} \right) \;.
  \label{Eq:pacc}
\end{equation}
For sufficiently long simulation times we can measure the equilibrium histogram in the radial ensemble
$n_w(R)$ which for any ensemble $w(R)$ allows us to determine the radial distribution function $g(R) = n_w(R) / w(R)$.

To speed up equilibration between the shells we optimize the statistical ensemble following the adaptive algorithm described in Ref.~\cite{OptimalEnsemble}. 
Initially we sample a flat histogram $n_w(R)=w(R)g(R)$ by setting the statistical weights to a fixed value $w(R) = 1/\tilde{g}(R)$, where a rough estimate of the radial distribution function $\tilde{g}(R)$ was obtained by a short run of the Wang-Landau algorithm \cite{WangLandau}. 
During the simulation batch we measure round-trips of the random walk along the radial reaction coordinate by adding a label to the random walker that indicates which of the two extremal radial distances $R_{\text{min}}$ or $R_{\text{max}}$ it has visited most recently.
We record two histograms, $n_w(R)$ and $n^+_w(R)$, where the histogram $n_w(R)$ is incremented in every Monte Carlo step, while the histogram $n^+_w(R)$ is increased only if the label indicates that with respect to the two extrema, $R_{\text{min}}$ and $R_{\text{max}}$, the random walker has visited $R_{\text{min}}$ most recently.
This allows to define a current from $R_{\text{min}}$ to $R_{\text{max}}$
\begin{equation}
  j = D(R) n_w(R) \frac{df}{dR} \;,
  \label{Eq:current}
\end{equation}
where $D(R)$ is the local diffusivity at distance $R$ and $f(R) = n^+_w(R)/n_w(R)$ is the fraction of walkers which have most recently visited $R_{\text{min}}$. Rearranging Eq.~(\ref{Eq:current}) gives a simple means to measure the local diffusivity 
\begin{equation}
  D(R) \sim  \left( n_w(R) \frac{df}{dR} \right)^{-1} \;.
\end{equation}
In Ref.~\cite{OptimalEnsemble} it was derived that in order to speed up equilibration along the reaction coordinate the current $j$ can be maximized by sampling an optimized histogram that is inversely proportional to the root of the local diffusivity
\begin{equation}
  n^{(opt)}_w(R) \sim \frac{1}{\sqrt{D(R)}} \;.
\end{equation}
To sample such a histogram the weights $w^{(opt)}(R)$ of an improved statistical ensemble are obtained as
\begin{equation}
  w^{(opt)}(R) = w(R) \sqrt{ \frac{1}{n_w(R)} \cdot \frac{df}{dR} } \;,
  \label{Eq:Feedback}
\end{equation}
which are then used to define the acceptance probabilities in Eq.~(\ref{Eq:pacc}) for a subsequent simulation batch. 
This feedback loop is iterated by simulating batches with repeatedly refined statistical weights $w(R)$ and increasing statistical accuracy. In our implementation we double the number of MC sweeps in subsequent batches, where one sweep corresponds to one attempted update per particle. The weights $w(R)$ are found to converge within four feedback iterations and a total of some $10^6-10^7$ MC sweeps.

\begin{figure}[t]
  \includegraphics[width=86mm]{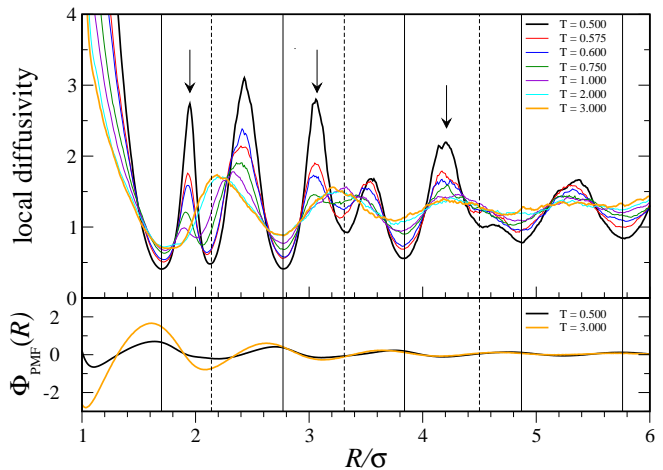}
  \caption{(Color online) 
                  Radial local diffusivity $D(R)$ for a dense Lennard-Jones fluid for varying temperatures. 
                  The free energy barriers between the shells of the liquid (see lower panel)
                  lead to a suppression of the local diffusivity. 
                  At low temperatures additional peaks (arrows) emerge revealing interstitial states 
                  between the shells of the liquid. 
                  The dashed lines indicate the interstitial barriers to which additional weight is shifted 
                  for the optimized ensemble/histogram. }
  \label{Fig:Diffusivity}
\end{figure}

{\em Simulations --}
We performed simulations for a 2D Lennard-Jones fluid with fixed density $\rho = 0.8$ at varying temperatures. We used a relatively small system of $N=144$ particles confined to a square box of length $13.42 \sigma$ with periodic boundary conditions. 
We simulated the pair-wise Lennard-Jones potential given in Eq.~(\ref{Eq:LJ}) without using a cut-off or shift. The continuous radial coordinate $R$ was binned into units of size $0.01\sigma$. In every update step a random displacement $\delta \leq 0.1\sigma$ in $x$- and $y$-directions was suggested for a single particle and accepted according to the acceptance probability in Eq.~(\ref{Eq:pacc}).  In order to calculate the radial distribution function $g(R)$ we simulated a broad histogram in the radial coordinate covering a range $[R_{\text{min}},R_{\text{max}}] \approx [0.95\sigma, 6.0\sigma]$. For each batch the local diffusivity is determined by recording two histograms, $n_w(R)$ and $n^+_w(R)$, and calculating the derivative $df/dR$ by a linear regression which then allows to perform the feedback step in Eq.~(\ref{Eq:Feedback}).  

\begin{figure}[t]
  \includegraphics[width=86mm]{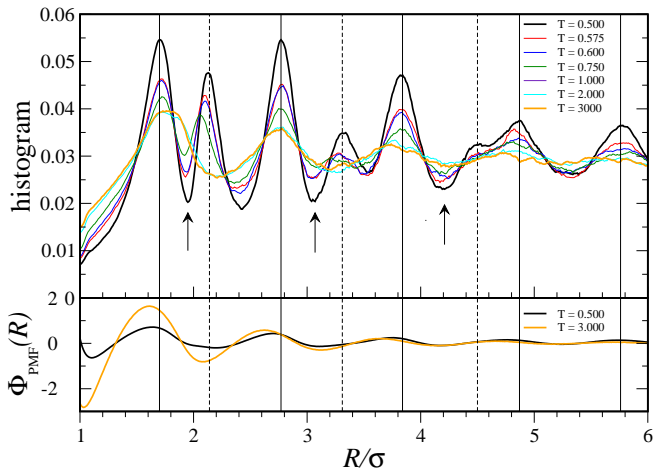}
  \caption{(Color online) 
                 Optimized histograms for a dense Lennard-Jones fluid after feedback of 
                 the local diffusivity for varying temperatures.
                 For the optimized ensemble peaks evolve at the free energy barriers 
                 between the shells of the liquid which proliferate at lower temperatures.
	        Additional peaks emerge at low temperatures (dashed lines)
	        revealing interstitial states (arrows) between the shells of the liquid. }
  \label{Fig:Histograms}
  \vspace{2mm}
\end{figure}

\begin{figure}[t]
  \includegraphics[width=86mm]{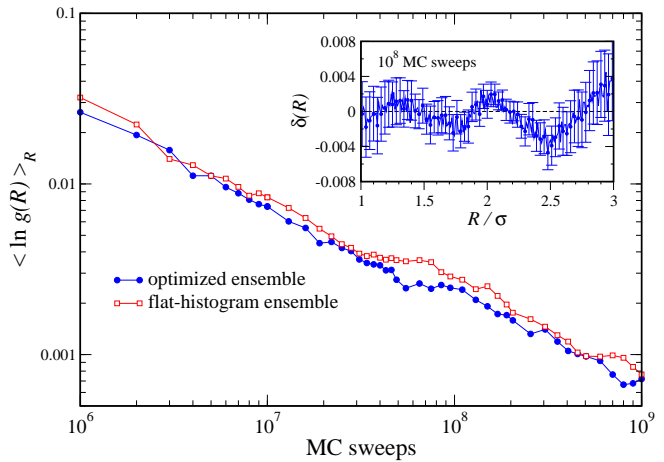}
  \caption{Average statistical error $\langle \Delta \ln g(R)\rangle_R$ of the computed radial distribution 
                 function for the optimized ensemble and flat-histogram sampling. 
                 Results are averaged over 16 independent Monte Carlo runs.
                 The inset shows the deviation of the radial distribution function obtained for  the optimized 
                 ensemble from a reference distribution function $\delta(R) = \ln g(R) - \ln g_{\text{ref}}(R)$ 
                 for 16 independent runs with $10^8$ MC sweeps. The reference distribution function was 
                 obtained from high statistics runs.}
  \label{Fig:Variance}
\end{figure}

In a first step we measure the local diffusivity $D(R)$ for the projected random walk along the radial distance $R$. The striking feature of the measured local diffusivity shown in Fig.~\ref{Fig:Diffusivity} is its strong variation along this reaction coordinate. Specifically, one observes for high temperatures $T>1$ that the local diffusivity has a maximum at the locations of the shells of the liquid (corresponding to the minima in the PMF) and is suppressed where the entropic barriers form between the shells. After feedback of the local diffusivity the sampled histogram is no longer flat, but peaks emerge at the locations of the barriers between the shells, see Fig.~\ref{Fig:Histograms}.
From the sampled histogram we have calculated the potential of mean force as $\Phi_{\text{PMF}}(R) =  \left( \ln w(R) - \ln n_w(R) \right)/\beta$ which is shown in Fig.~\ref{Fig:PMF}. 
We find that the numerical accuracy of the estimated radial distribution function and PMF is moderately improved with an average speedup of $50-100\%$ 
by sampling the optimized histogram instead of a flat histogram as shown in Fig.~\ref{Fig:Variance}.

\begin{figure}[b]
  \includegraphics[width=\columnwidth]{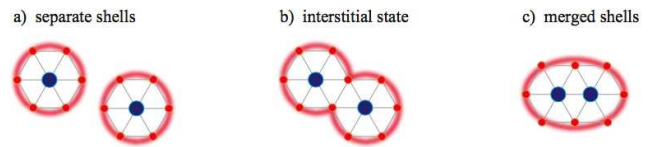}
  \caption{Shell structures in a dense Lennard-Jones fluid. 
                  a) Two particles at large distance with separate shells and a total of 12 neighbors.
                  b) An interstitial state forms as the shells around two particles merge with a characteristic
                       distance of $R=\sqrt{3}\sqrt[6]{2}\sigma \approx 1.94\sigma$ and a total of 10 neighbors. 
                  c) Pair of particles confined to the first shell with a characteristic distance of 
                       $R=\sqrt[6]{2}\sigma \approx 1.12\sigma$ and a total of 8 neighbors. }
  \label{Fig:InterstitialState}
\end{figure}

{\em Interstitial states --}
As the temperature is further lowered to $T \leq 1$ and the system is driven towards vapor-liquid coexistence, see Fig.~\ref{Fig:PMF}, additional features emerge in both the measured local diffusivity and the optimized histogram, indicated by the arrows in Figs.~\ref{Fig:Diffusivity} and \ref{Fig:Histograms}, that point to the formation of additional structures.
In the local diffusivity a sharp additional peak emerges between the first and second shell at radial distance $R \approx 1.94\sigma$ as shown in Fig.~\ref{Fig:Diffusivity}. The close proximity of this peak to $R=\sqrt{3}\sqrt[6]{2}\sigma$ reveals that an interstitial state becomes stable at low temperatures which is formed when the first shells around the two particles merge as illustrated in Fig.~\ref{Fig:InterstitialState} b).
Similarly, additional peaks between the higher shells point to interstitial states that are formed when the second/third shells around two particles merge. 

\begin{figure}[t]
  \includegraphics[width=86mm]{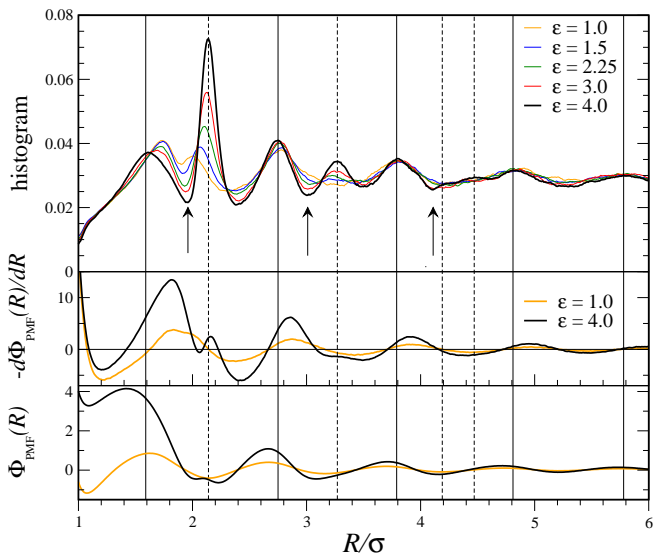}
  \caption{(Color online) 
                 Optimized histograms for a particle-solvent model with varying interaction strength 
                  $\epsilon$.
                 The solid lines indicate barriers between the shells of the liquid, the dashed
                  lines refer to interstitial barriers, and the arrows indicate interstitial states.
                  The second panel shows the average force, that is the derivative of the potential of mean force 
                  shown in the third panel.}
  \label{Fig:Histograms_epsilon}
\end{figure}

The distinct signature of the interstitial states in these plots unveils the potential use of the optimized ensemble Monte Carlo method to identify interstitial states in the simulation of more complex systems such as fluids undergoing a liquid-liquid phase transition \cite{Lee:01}, amorphous states \cite{Martonak} or biological systems. For instance, recent neutron diffraction experiments \cite{Klotz:02} with high-density amorphous (HDA) ice \cite{SupercooledWater} revealed an interstitial molecular state which distinguishes HDA from low-density amorphous (LDA) ice. Molecular dynamics simulations \cite{Martonak} confirmed this scenario by measuring the radial oxygen-oxygen distribution function $g(R)$. While the interstitial state can be easily tracked in the pair distribution function for the high-density phase, the signature quickly faints as the pressure is reduced and a more sensitive measure such as the optimized histogram may help to identify the nature of the transition. 
For comparison we have indicated the location of the interstitial states also in the potential of mean force and its derivative, the average force, in Fig.~\ref{Fig:Histograms_epsilon}. In contrast to the optimized histogram only a weak feature in the average force points to the interstitial state between the first and second shell which was also pointed out in Ref.~\cite{AIM}. 

{\em Particle-solvent model --}
As another application of the optimized ensemble method we have studied a simple particle solvent model where we vary the strength $\epsilon$ given in Eq.~(\ref{Eq:LJ}) of an attractive interaction between a pair of particles and a solvent. Within the solvent we assume uniform coupling $\epsilon = 1$. Varying the interaction strength $\epsilon$ between the particles and the solvent is equivalent to a local renormalization of the temperature. As the interaction strength is increased at a fixed temperature $T=1$ shells of the solvent form around the particles as illustrated in Fig.~\ref{Fig:InterstitialState} a). When the two particles approach each other the break-up of these shells is protected by a free energy barrier.   
Optimizing the radial ensemble statistical weight is shifted towards this barrier and a peak in the histogram emerges at the location of the barrier as shown in Fig.~\ref{Fig:Histograms_epsilon}.
For increasing interaction strength $\epsilon$ this peak at $R\approx 2.14\sigma$ further proliferates and becomes the dominant feature in the histogram. Due to the local character of the temperature renormalization upon increasing $\epsilon$ the peaks at the locations of interstitial barriers between higher shells are only weakly amplified.


{\em Conclusions --}
In conclusion, by simulating a dense Lennard-Jones liquid we demonstrated that the optimized broad-histogram ensemble for a system with multiple entropic barriers can be efficiently determined by the recently developed adaptive ensemble optimization technique \cite{OptimalEnsemble}. 
The algorithm allows to systematically shift statistical weight towards the barriers between the shells of the liquid by measuring the local diffusivity along a continuous radial reaction coordinate.
Distinct features in the measured local diffusivity and optimized histogram point to interstitial states and allow to significantly increase the sensitivity to detect such states compared to other approaches based on measurements of the radial distribution function or average force.
The ensemble optimization method thereby provides a simple means to detect interstitial states that should be applicable to more complex systems.


{\em Acknowledgments --}
We thank R.~H.~Swendsen and R. Marto\v{n}\'ak for stimulating discussions. 
ST acknowledges support by the Swiss National Science Foundation. 
MT acknowledges support by the Aspen Center for Physics.


\end{document}